\documentclass{ieeeaccess}
\usepackage{cite}
\usepackage{amsmath,amssymb,amsfonts}
\usepackage{algorithmic}
\usepackage{graphicx}
\usepackage{textcomp}
\usepackage{optidef}

\usepackage{booktabs} %table formatting
\usepackage{color, colortbl}  % table shading
\usepackage{pifont} % tick and x marks
\usepackage{mathtools} %bra-ket
\usepackage{xurl,lineno,microtype}
\usepackage{pythonhighlight}
\usepackage{bm}
\makeatletter
\AtBeginDocument{\DeclareMathVersion{bold}
\SetSymbolFont{operators}{bold}{T1}{times}{b}{n}
\SetSymbolFont{NewLetters}{bold}{T1}{times}{b}{it}
\SetMathAlphabet{\mathrm}{bold}{T1}{times}{b}{n}
\SetMathAlphabet{\mathit}{bold}{T1}{times}{b}{it}
\SetMathAlphabet{\mathbf}{bold}{T1}{times}{b}{n}
\SetMathAlphabet{\mathtt}{bold}{OT1}{pcr}{b}{n}
\SetSymbolFont{symbols}{bold}{OMS}{cmsy}{b}{n}
\renewcommand\boldmath{\@nomath\boldmath\mathversion{bold}}}
\makeatother

\def\BibTeX{{\rm B\kern-.05em{\sc i\kern-.025em b}\kern-.08em
    T\kern-.1667em\lower.7ex\hbox{E}\kern-.125emX}}

\definecolor{GREEN}{RGB}{0, 0, 0}
\newcommand{\hltext}[1]{\textcolor{GREEN}{#1}}
\renewcommand{\theenumi}{\Alph{enumi}}

\DeclarePairedDelimiterX\braket[2]{\langle}{\rangle}{#1\,\delimsize\vert\,\mathopen{}#2}

\begin{document}
\history{ }
\doi{ }

\title{D-Wave's Nonlinear-Program Hybrid Solver: Description and Performance Analysis}
\author{\uppercase{Eneko Osaba}\authorrefmark{1} AND \uppercase{Pablo Miranda-Rodriguez}\authorrefmark{1}}

\address[1]{TECNALIA, Basque Research and Technology Alliance (BRTA), 48160 Derio, Spain}

\tfootnote{This work was supported by the Basque Government through ELKARTEK Program under Grants KK-2024/00105 (KUBIT project) and KK-2024/00024 (AURRERA project).}

\markboth
{Osaba and Miranda-Rodriguez: D-Wave's Nonlinear-Program Hybrid Solver}
{Osaba and Miranda-Rodriguez: D-Wave's Nonlinear-Program Hybrid Solver}

\corresp{Corresponding author: Eneko Osaba (email: eneko.osaba@tecnalia.com).}

\begin{abstract}
The development of advanced quantum-classical algorithms is among the most prominent strategies in quantum computing. Numerous hybrid solvers have been introduced recently. Many of these methods are created ad hoc to address specific use cases. However, several well-established schemes are frequently utilized to address optimization problems. In this context, D-Wave launched the \textit{Hybrid Solver Service} in 2020, offering a portfolio of methods designed to accelerate time-to-solution for users aiming to optimize performance and operational processes. Recently, a new technique has been added to this portfolio: the \textit{Nonlinear-Program Hybrid Solver}. This paper describes this solver and evaluates its performance through a benchmark of 45 instances across three combinatorial optimization problems: the Traveling Salesman Problem, the Knapsack Problem, and the Maximum Cut Problem. To facilitate the use of this relatively unexplored solver, we provide details of the implementation used to solve these three optimization problems.
\end{abstract}


\titlepgskip=-21pt

\maketitle

%-------------------------
\section{Introduction}
\label{sec:introduction}
%-------------------------

%EL MOMENTUM DE LA QUANTICA
\PARstart{T}{he} emergence of quantum technology is expected to have a significant impact on a number of industries. The field of Quantum Computing (QC), which leverages the principles of quantum mechanics to process information, is constantly making advances to connect quantum processing with practical use cases. Thus, QC has advanced significantly in recent years, mainly due to the rapid development of technology and advancements in \hltext{its democratization, understanding it as the process of making QC accessible to a broader spectrum of individuals and communities \cite{seskir2023democratization,democrat}}. As a result, QC has facilitated the development of various proofs of concept across multiple sectors as finance \cite{herman2023quantum}, energy \cite{joseph2023quantum}, and logistics \cite{osaba2022systematic}.

\hltext{Historically, the idea of a quantum computer can be traced back to the works of Benioff \cite{Benioff1980} and Feynman\cite{Feynman1982}, where they proposed a quantum mechanical implementation of a Turing machine and the idea of simulating a quantum system with another quantum system, respectively.}
Currently, there are two types of real quantum devices that we can differentiate between: gate-based quantum computers and quantum annealers. A gate-based system, on the one hand, employs basic quantum circuit operations on qubits, which are akin to the classical operations on regular bits and may be combined in any order to create algorithms. This version is often referred to as \textit{a universal quantum computer}. On the other hand, a quantum annealer \hltext{is based on adiabatic computation, wherein an initial, easily prepared Hamiltonian is gradually and continuously evolved from its ground state to the ground state of a final, problem-specific Hamiltonian. If the evolution is slow enough, the adiabatic theorem guarantees that the system remains in the ground state during the whole computation. In quantum annealers, the adiabatic theorem is intentionally relaxed, allowing the system to evolve more rapidly than the adiabatic limit would dictate. As a result, transitions to higher energy states usually occur during the evolution; however, alternative methods to achieve adiabaticity have been proposed \cite{Takahashi_2017,ferreirovélez2024shortcutsadiabaticvariationalalgorithms}. Although this model of computation is also universal \cite{PhysRevLett.99.070502}, D-Wave quantum annealer is based on an Ising Hamiltonian, restricting the types of problems that can be run on the system. This type of quantum annealer is, however, well suited to solve combinatorial optimization problems \cite{yang2023survey}.}

%LIMITACIONES
Despite all the progress made in the field, quantum computers are still in their infancy in contrast to classical computers, which have been developed for decades and are therefore highly advanced. Thus, quantum devices are currently unable to efficiently solve real-world problems mainly because of the small number of qubits and their unstable nature. \hltext{Phenomena such as decoherence time, noise and information loss, in absence of error correction protocols, impact in the performance of the computation. Besides, there are other roadblocks such as quantum gate fidelity and gate noise}. At this time, the number of qubits for both universal quantum computers and quantum annealers is around three orders of magnitude (thousands of qubits) \cite{IBM1000,Advantadge2}. However, this number needs to be much greater in order for the technology to become truly useful for real industrial use cases. In any case, there are factors other than the qubit count that influence the practical capabilities of quantum computers, \hltext{such as the connectivity topology or the previously mentioned operational fidelity and decoherence time.}

As a result of this situation, recent advancements have emerged during the \textit{noisy intermediate-scale quantum} (NISQ, \cite{preskill2018quantum}) era, a period characterized by quantum computers' limitations in efficiently handling problems, even those of small to medium size. As it turns out, both universal quantum devices and annealers suffer from these limitations. 

%LA COMPUTACIÓN HÍBRIDA ES EL FUTURO
As a consequence, the entire community is striving to come up with mechanisms to deal with the present limitations and capitalize on the promise that QC has to offer. The design of advanced hybrid algorithms, which combine the advantages of both computing paradigms, are among the most popular strategies \cite{endo2021hybrid}. Arguably, hybrid quantum computing represents the immediate future of this area. This is so because the adoption of quantum techniques to address real-world use cases is heavily reliant on hardware capabilities. In this regard, just as QC should not be thought of as a direct replacement for conventional computing, it would also be a mistake to view quantum-classical hybrid computing as merely a temporary fix to minimize the limitations of NISQ-era systems. As stated in works such as \cite{callison2022hybrid}, hybrid algorithms will be influential well beyond the NISQ-era and even into full fault tolerance, with quantum computers enhancing the capabilities of already powerful classical processors by carrying out certain specialized tasks. The challenge here is in determining how to integrate classical and quantum computing to create a synergy that surpasses the performance of purely classical approaches. 

\hltext{Besides, hybrid algorithms have its own limitations. For example, since quantum and classical hardware are often physically separated and they need to share information, latency is introduced in the workflow \cite{latency}. Another roadblock is the possibility of barren plateaus in the optimization landscape, making it difficult for classical optimizers to converge \cite{larocca2024reviewbarrenplateausvariational}, \cite{Ragone_2024}. Also, when the problem at hand exceeds the computational capacity of the quantum processing unit (QPU), determining which subproblems to allocate to the quantum component of the algorithm is a non-trivial task \cite{Booth2017}, \cite{VERMA2022100594}. The choice of methodology is not always straightforward and often depends on the specific characteristics of the problem and the quantum hardware's limitations, for example its topology.}

%ES COMPLEJO IMPLEMENTAR ALGORITMOS CUANTICOS
It is also crucial to note that the design and implementation of quantum algorithms could necessitate a high degree of subject matter expertise. This intricacy could be a hindrance for researchers who lack a sufficient background in disciplines like physics or quantum mechanics. In order to remove that barrier and make QC more accessible, various frameworks and programming languages are proposed, such as Silq \cite{bichsel2020silq}, Eclipse Qrisp \cite{seidel2024qrisp}, or Qiskit \cite{javadi2024quantum}. The availability of such frameworks and languages helps to foster the building of a multidisciplinary community focused on QC and helps the field to progress toward new horizons \cite{villar2023hybrid}.

%%%CONTEXTO DE ALGORITMOS HÍBRIDOS NO SOLO DE D-WAVE (COMENTARIO HECHO PARA EL REVISOR 2
\hltext{In this article, we focus on the hybrid algorithms and frameworks proposed by the Canadian company D-Wave Systems. Anyway, it is worth noting that significant efforts have been made from various fronts in the design and implementation of both hybrid techniques and platforms. Probably the most well-known and widely studied hybrid resolution schemes are the Variational Quantum Algorithms (VQA, \cite{cerezo2021variational})), oriented towards gate-based quantum computing. The most representative examples of VQA are the Quantum Approximate Optimization Algorithm (QAOA, \cite{farhi2014quantum}) and the Variational Quantum Eigensolver (VQE, \cite{tilly2022variational}). In this context, it is also interesting to highlight commercial approaches such as the so-called \texttt{HybridSolver} developed by the German company Quantagonia\footnote{https://www.quantagonia.com/hybridsolver}. Regarding frameworks and platforms specifically created to facilitate the design, implementation, and execution of hybrid algorithms, notable examples include NVIDIA's CUDA-Q\footnote{https://developer.nvidia.com/cuda-q}, IBM Qiskit Runtime\footnote{https://docs.quantum.ibm.com/api/qiskit-ibm-runtime/runtime\_service} and Google's Cirq\footnote{https://quantumai.google/cirq}.}

%D-WAVE Y EL HYBRID FRAMEWORK
Focusing our attention on D-Wave, and as part of the strategy to bring QC to a wider audience, \textit{D-Wave-Hybrid-Framework} was published in 2018, which, in the words of its creators, is ``\textit{a general, minimal Python framework to build hybrid asynchronous decomposition samplers for QUBO problems}"\footnote{\url{https://github.com/dwavesystems/dwave-hybrid}}. This framework is appropriate for ``\textit{developing hybrid approaches to combining quantum and classical compute resources}". \

%INTRODUCCION HSS
Some further efforts were made by D-Wave to bring hybrid solvers to industry. Thus, in 2020, \textit{D-Wave's Hybrid Solver Service} (HSS, \cite{HSS}) was launched, which consists of a portfolio of hybrid solvers that leverage quantum and classical computation to tackle large and/or real-world optimization problems. HSS is tailored for researchers and practitioners aiming to streamline the code development process. Consequently, all solvers in the portfolio are designed to enhance time-to-solution, aiding users in optimizing performance and operational workflows.

Until June 2024, the HSS included three solvers for solving three different problem types \cite{leapCQM}: the binary quadratic model (BQM) solver, \texttt{BQM-Hybrid}, for problems defined using binary variables; the discrete quadratic model (DQM) technique, \texttt{DQM-Hybrid}, for problems defined on discrete values; and the constrained quadratic model (CQM) method, \texttt{CQM-Hybrid}, which can deal with problems defined on binary, integer, and even continuous variables. Recently, a new solver to be added to the portfolio has been unveiled: the \textit{Nonlinear-Program Hybrid Solver}, or \texttt{NL-Hybrid}. 

As we have summarized systematically in Table \ref{tab:survey}, much research has been carried out in recent years around \texttt{BQM-Hybrid}, \texttt{DQM-Hybrid}, and \texttt{CQM-Hybrid}. \hltext{Among these methods, \texttt{BQM-Hybrid} is the one most frequently used by the scientific community. This is because it was the first to be implemented and its usage is similar to the D-Wave QPU. It is important to highlight that, precisely for this reason, the \texttt{BQM-Hybrid} is not well-suited for solving real-world problems with many constraints. Consequently, there are few studies where this method has been used to solve problems with realistic aspects \cite{nourbakhsh2022exploring,fox2021mrna,fox2022rna,iturrospe2021optimizing}. With all this, the \texttt{BQM-Hybrid} has been primarily used as a benchmarking algorithm \cite{colucci2023power,makarov2023optimization,makarov2024quantum,dinh2023efficient,schworm2023solving,phillipson2021portfolio,mugel2022dynamic,certo2022comparing}. That is, as a method employed within an experimentation to measure the performance of other methods. Moreover, \texttt{BQM-Hybrid} has also been used to solve academic problems, such as Tail Assignment Problem \cite{martins2021qubo}, Vehicle Routing Problem \cite{tambunan2023quantum} and the Set Packing Problem \cite{venere2023design}.}

\hltext{Focusing on the \texttt{CQM-Hybrid}, we see how this advanced solver has a clear orientation towards solving real-world problems. The \texttt{CQM-Hybrid} has been used in a significant number of works to address complex problems with a high number of constraints, and it has been employed in different contexts: to solve a complete problem \cite{v2023hybrid,romero2023solving,benson2023cqm,colucci2023power,schworm2023solving,nau2022hybrid,nayak2022quantum}; as part of a complex resolution pipeline where it is called a single time \cite{o2023quantum} or iteratively \cite{osaba2024solving}; or to implement mechanisms to improve the performance of existing classical algorithms \cite{osaba2024exploring}. Furthermore, the \texttt{CQM-Hybrid} has been scarcely used as a benchmarking algorithm \cite{certo2022comparing,tripathy2022comparative}.}

\hltext{Regarding the \texttt{DQM-Hybrid}, it is the HSS method that has received the least attention from the community. This is mainly because it is less flexible and powerful compared to \texttt{CQM-Hybrid}, leading researchers to clearly favor the latter. Thus, the \texttt{DQM-Hybrid} has been used in only a handful of papers, either to act as a benchmarking method \cite{schworm2023solving} or to solve clustering and community detection problems \cite{fernandez2021community,wierzbinski2023community,jhaveri2023quantum}.}

\begin{table}[t]
  \centering
  \caption{Brief survey on recent practical applications of HSS solvers, classified by the field of knowledge on which the research is focused. }
  \label{tab:survey}
    \resizebox{1.0\columnwidth}{!}{
        \begin{tabular}{lccc}
            \toprule
            \textbf{Field} & \texttt{BQM-Hybrid} & \texttt{DQM-Hybrid} & \texttt{CQM-Hybrid}\\
            \midrule
            \textbf{Logistics} & \cite{martins2021qubo,tambunan2023quantum} & & \cite{osaba2024solving,romero2023solving}\\
            \textbf{Medicine} & \cite{nourbakhsh2022exploring,fox2021mrna,fox2022rna} & \cite{wierzbinski2023community} & \cite{benson2023cqm}\\
            \textbf{Energy} & \cite{colucci2023power,fernandez2021community} & \cite{fernandez2021community} & \cite{colucci2023power,o2023quantum}\\
            \textbf{Industry} & \cite{makarov2023optimization,makarov2024quantum,dinh2023efficient,schworm2023solving} & \cite{schworm2023solving} & \cite{dinh2023efficient,nau2022hybrid,nayak2022quantum,v2023hybrid,schworm2023solving,osaba2024exploring}\\
            \textbf{Finance} & \cite{phillipson2021portfolio,mugel2022dynamic,certo2022comparing} & & \cite{certo2022comparing,tripathy2022comparative}\\
            \textbf{Others} & \cite{pasetto2022quantum,iturrospe2021optimizing,venere2023design} & \cite{jhaveri2023quantum} & \\
            \bottomrule
        \end{tabular}
    }
\end{table}

%INTRODUCIMOS NUESTRO TRABAJO
Despite this abundant scientific activity, no work has yet been published on \texttt{NL-Hybrid}. Motivated by the lack of existing research, this paper focuses on describing the newly introduced \texttt{NL-Hybrid} solver. Furthermore, we will conduct an experiment to analyze the performance of this new solver in comparison with \texttt{BQM-Hybrid}, \texttt{CQM-Hybrid}, and D-Wave's QPU \texttt{Advantage\_system6.4}. For these tests, we used a benchmark composed of 45 instances, equally distributed across three combinatorial optimization problems: the Traveling Salesman Problem (TSP, \cite{lin1965computer}), the Knapsack Problem (KP, \cite{martello1987algorithms}), and the Maximum Cut Problem (MCP, \cite{hadlock1975finding}). \hltext{We have chosen these problems because:} 

\begin{itemize}
	\item \hltext{They have been extensively used for benchmarking purposes in QC-oriented research \cite{bozejko2024optimal,wang2018quantum,osaba2022systematic}.}
	\item \hltext{They are appropriate for formulation and use in the solvers considered in this study. Furthermore, TSP and KP are well suited to be solved with \texttt{NL-Hybrid} newly introduced variables, while MCP can be easily formulated with traditional binary variables.}
	\item \hltext{Their complexity for being solved by QC-based methods has been previously demonstrated \cite{van2021quantum,pecci2024beyond,qian2023comparative}.}
\end{itemize}

Finally, to facilitate the use of this still unexplored solver, we will share details of the implementation to solve these three optimization problems.

%POR PETICIÓN DEL REVISOR 3, AQUÍ NOS DISTINGUIMOS DE NUESTRO TRABAJO ANTERIOR
\hltext{This research complements previous work on hybrid algorithms carried out by the authors of this study. In works such as \cite{osaba2021hybrid,osaba2021focusing}, the authors presented ad hoc implemented hybrid algorithms to solve the TSP and its asymmetric variant. Furthermore, in \cite{osaba2024exploring,osaba2024solving,romero2023solving,v2023hybrid}, the authors used the \texttt{CQM-Hybrid} to tackle real-world problems in fields such as logistics and industry. Finally, the authors have also conducted research focused on benchmarking different hybrid algorithms, with representative works such as \cite{osaba2024eclipse,osaba2024hybrid,osaba2023qoptlib}. This paper stands out from all these papers, being the first to work with and experiment on the newly unveiled \texttt{NL-Hybrid}}.

The rest of this article is structured as follows. In Section \ref{sec:NL}, we provide a brief background related to HSS and describe the main aspects of \texttt{NL-Hybrid}. Section \ref{sec:implementation} focuses on introducing the implementation details. Section \ref{sec:expres} details the experimentation carried out. The paper ends with Section~\ref{sec:conclusion}, which draws conclusions and outlines potential avenues for future research.

%-------------------------
\section{Nonlinear-Program Hybrid Solver}
\label{sec:NL}
%-------------------------
We divide this section on the \texttt{NL-Hybrid} into two parts. First, in Section \ref{sec:desc}, we give an overview of the method. Then, in Section \ref{sec:work}, we take a look at its structure and workflow.

\subsection{Overview of NL-Hybrid}
\label{sec:desc}

In recent years, the community has proposed a plethora of hybrid solvers. Many of these techniques are ad hoc implementations to tackle specific problems. Usually, researchers consider a number of factors when designing their hybrid solvers, such as \textit{i)} the specifics of the problem to be solved; \textit{ii)} the limitations and characteristics of the quantum device to be used; and/or \textit{iii)} the knowledge and intuition of the developer. In many cases, the researcher's knowledge in fields such as artificial intelligence or optimization is crucial \cite{precup2020experiment}.

However, there are some well-established methods that the community routinely uses, such as the aforementioned QAOA and VQE, or QBSolv \cite{qbsolv}. Another interesting and recognizable scheme is \textit{Kerberos} \cite{malviya2023logistics}, which is a concretization of the above-described \textit{D-Wave Hybrid Framework}. More specifically, \textit{Kerberos} is a reference hybrid workflow comprised of three methods iteratively running in parallel: two classical methods, Simulated Annealing and Tabu Search, and a quantum one that uses the QPU of D-Wave.

As mentioned beforehand, D-Wave's main motivation for creating HSS is to introduce a portfolio of hybrid techniques that lighten the method-implementation phase. In this way, HSS is made up of a set of four ready-to-use methods that target different categories of input and use cases. The last of the methods included in HSS is the \texttt{NL-Hybrid}, which is the one we will focus on in this article and which represents a breakthrough in the implementation of hybrid algorithms.

Firstly, \texttt{NL-Hybrid} stands out because it allows for the definition of variables in \hltext{other additional} formats than those considered in the methods previously included in HSS. More specifically, \texttt{NL-Hybrid} excels with decision variables that embody common logic, such as ordering permutations or subsets of options. For instance, in routing problems like the TSP, a permutation of variables indicates the sequence in which nodes are visited. Similarly, in the KP, the variables representing items to be stored can be categorized into two distinct groups: packed and unpacked.

In this way, in addition to allowing variables defined as binary and integer values, \texttt{NL-Hybrid} permits the definition of the following types of decision variables:

\begin{itemize}
    \item \texttt{list(number\_variables)}: The solver can use a \texttt{list} as the decision variable to optimize, this being an ordered permutation of size \texttt{number\_variables} describing a possible itinerary.
    
    \item \texttt{set(number\_variables)}: The decision variable can be a \texttt{set}, being this a subset of an array of size \texttt{number\_variables}, representing possible items included in a knapsack.
   
    \item \texttt{disjoint\_list(n\_variables,n\_lists)}: The solver can employ a \texttt{disjoint\_list} as the decision variable, which divides a set of \texttt{n\_variables} into \texttt{n\_lists} disjoint ordered partitions, each representing a permutation of variables. This encoding is appropriate for complex logistic problems such as the Vehicle Routing Problem. There is a variant of this variable, called \texttt{disjoint\_bit\_sets}, where the order of the produced partitions is not semantically meaningful.
\end{itemize}
    
It is worth mentioning at this point that in the field of optimization, whether by means of classical or quantum systems, \hltext{the performance of a solver is closely tied to its capacity to explore and exploit the whole \textit{solution space}. That is, the larger the \textit{search space}, the higher the probability of reaching better solutions, understanding \textit{search space} as the region of the \textit{solution space} that the algorithm can access. However, there is the downside that larger spaces are usually computationally expensive to explore and exploit. Therefore, employing decision variables that act as implicit constraints is an effective way to reduce both \textit{search} and \textit{solution spaces} and thus, the running time to find a solution}. For instance, using \texttt{list(number\_variables)} to represent a canonical TSP implicitly ensures that no nodes are visited more than once along the route \hltext{and also that each node is visited}. \hltext{This results in the absence of infeasible solutions within the \textit{solution space}, thereby ensuring that this space comprises solely feasible solutions to the problem. This characteristic is not observed with binary encoding}. For this reason, the use of the aforementioned decision variables is a significant advantage for \texttt{NL-Hybrid}.

Finally, \texttt{NL-Hybrid} natively permits \hltext{nonlinear (linear, quadratic, and higher order)} inequality and equality constraints, expressed even arithmetically. This aspect represents a significant contribution compared to other well-established hybrid solvers. \hltext{For comparison, \texttt{BQM-Hybrid} accepts linear soft constrains as penalty models, while \texttt{CQM-Hybrid} works with linear and quadratic constrains natively, although it is also possible to implement them as soft constraints.}

To conclude this section, it is important to note that the existence of \texttt{NL-Hybrid} does not imply the complete deprecation of previously existing methods in HSS. Depending on the characteristics of the problem and the decision variables used, \texttt{NL-Hybrid} may not always be the most efficient algorithm. As will be demonstrated in later sections, \texttt{CQM-Hybrid} or \texttt{BQM-Hybrid} might be more suitable for problems primarily composed of binary variables.

\subsection{Structure and Workflow of NL-Hybrid}
\label{sec:work}

\begin{figure}[t]
    \centering
    \includegraphics[width=0.85\linewidth]{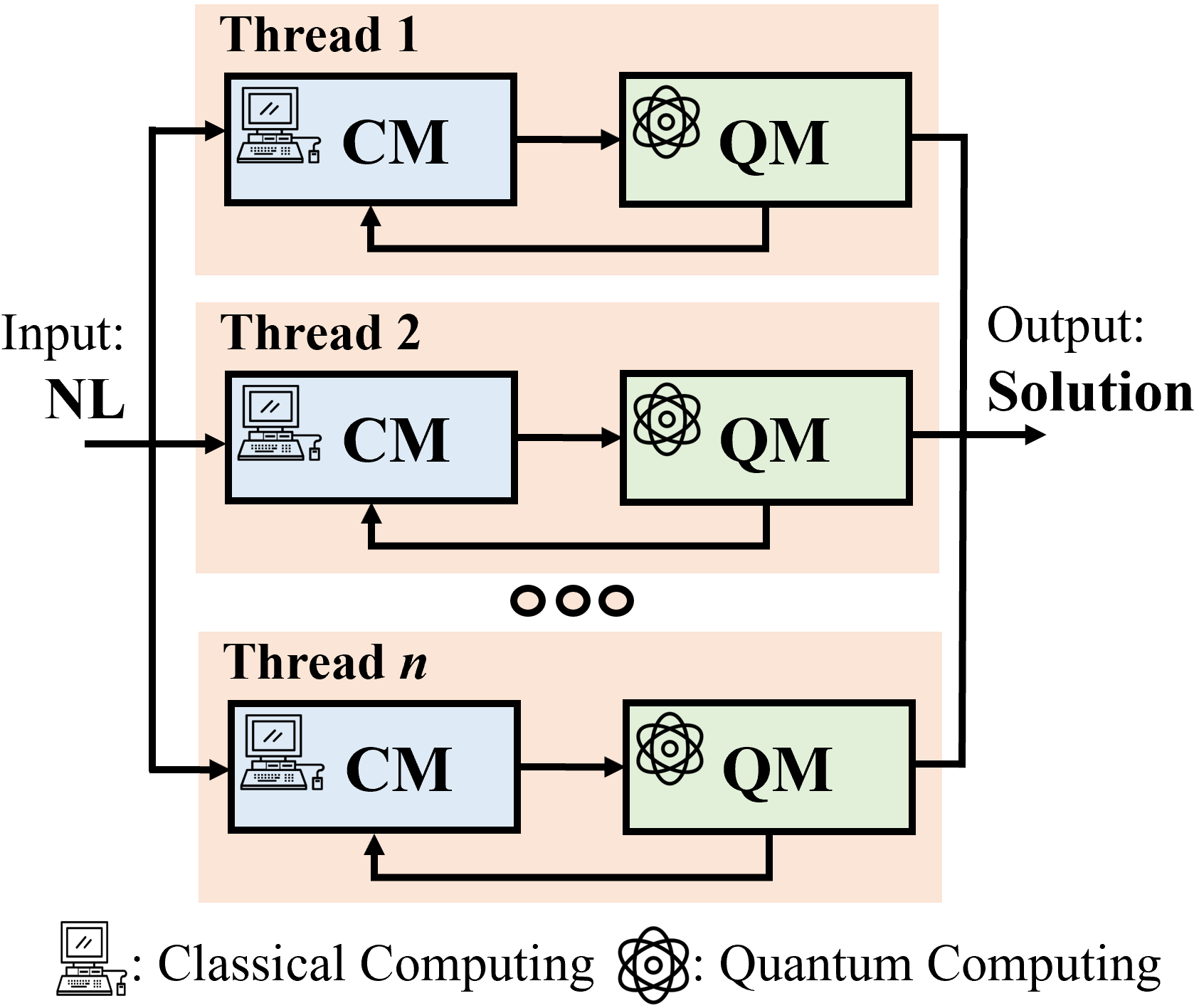}
    \caption{Structure of NL-Hybrid solver. CM = Classical Heuristic Module. QM = Quantum Module.}
    \label{fig:hss}
\end{figure}

Being part of HSS, \texttt{NL-Hybrid} has the same structure as the other methods within the portfolio. This structure, which is depicted in Figure \ref{fig:hss}, is divided into three distinct phases:

\begin{enumerate}
    \item First, the NL formulation of the problem is introduced as input into a classical front end. In this preliminary phase, the solver creates a predetermined number of equally structured branches.

    \item Secondly, each created thread is executed in parallel on a set of Amazon Web Services (AWS) CPUs and/or GPUs. Each branch is composed of a Classical Heuristic Module (CM) and a Quantum Module (QM). The CM is in charge of exploring the problem-solution space using traditional heuristics. During this exploration, the CM formulates different quantum queries, which are executed by the QM, and which are partial representations of the problem that are accommodated to the QPU capacity. The solutions provided by the QPU are employed to guide the CM toward promising areas of the solution search space. Furthermore, QM can even improve the solutions found by the CM. \texttt{NL-Hybrid} resorts to the latest D-Wave quantum device to execute the quantum queries. At the time of this writing, the system used was the \texttt{Advantage\_system6.4}, which is made up of 5616 qubits organized in a Pegasus topology. 

    \item Finally, after a predefined time limit $T$, all generated branches stop their execution and return their solution to the front end. Then, \texttt{NL-Hybrid} forwards the best solution found among all the threads. It should be noted that CM and QM communicate asynchronously, ensuring that latency in a particular branch does not hinder the overall progress of the \texttt{NL-Hybrid} solver.

\end{enumerate}

Some of the benefits of using \texttt{NL-Hybrid} over ad hoc generated methods or other widely recognized solvers are:

\begin{itemize}
    \item \texttt{NL-Hybrid} is built to manage low-level operational specifics, eliminating the need for users to have any expertise in properly parameterizing the QPU.

    \item \texttt{NL-Hybrid} accepts inputs that are much larger than those of other solvers focused on solving problems in QUBO format and even larger than those of the rest of the solvers within HSS. \texttt{NL-Hybrid} is intended to take advantage of the QPU's capability to quickly find promising solutions, expanding this property to a wider range of input types and sizes than would otherwise be feasible.

    \item \texttt{NL-Hybrid} provides a user-friendly use of quantum resources, allowing the user to model a problem in an intuitive way. This is an advantage in comparison to QUBO, which is the native formulation for QPUs, mainly because translating a problem to this binary formulation is often a challenging task \cite{de2024optimized}. In fact, inefficient translation can critically affect the performance of the solver.   
\end{itemize}

\hltext{Lastly, it should be noted that the \texttt{NL-Hybrid} solver is proprietary. Consequently, further technical details are not available to the general public. For additional information on this method, we refer interested readers to the D-Wave report on the HSS portfolio \cite{HSS}}.

%-------------------------
\section{Implementation Details}
\label{sec:implementation}
%-------------------------

The experiments conducted in this study focus on three combinatorial optimization problems: TSP, KP, and MCP. Due to its incipient nature, there is little information regarding \texttt{NL-Hybrid}. Given this lack of documentation, we provide several key implementation details on how to tackle the above-mentioned problems by means of \texttt{NL-Hybrid}. This allows the reader to understand how intuitive the problem design is.

It should be noted that, while the implementation of MCP was done from scratch, those for TSP and KP are slight adaptations of the open-source code published by D-Wave\footnote{\url{https://github.com/dwavesystems/dwave-optimization/blob/main/dwave/optimization/generators.py}}.

First, in order for the problem to be solved by \texttt{NL-Hybrid}, it must be defined using a special entity dedicated for this purpose, called \texttt{Model}. Once this model is initialized, the definition of a problem includes the following four steps:

\begin{itemize}
    \item Defining the decision variables.
    \item Entering the information necessary to describe the problem as \texttt{constants} (if needed). 
    \item Defining the problem constraints (if any).
    \item Formulating the objective function.
\end{itemize}

The rest of the section is divided into four subsections. The first three are devoted to each of the problems to be addressed, while the last is devoted to the execution of the \texttt{NL-Hybrid}.

%-------------------------
\subsection{Traveling Salesman Problem}
\label{sec:TSP}
%-------------------------

The TSP is a classical routing problem that can be represented as a complete graph $G=(V,A)$ of size $N$, where $V$ illustrates the set of nodes and $A$ the set of edges linking every pair of nodes in $V$. Furthermore, a matrix $C$ of size $NxN$ contains the costs $c_{ij}$ associated with traveling from node $i$ to node $j$.

The TSP is a problem that is particularly suitable for being solved by \texttt{NL-Hybrid}, since the decision variables can be defined using the above-described \texttt{list(number\_variables)} type. Thus, \texttt{list(N)} represents an ordered permutation of the nodes in $V$. It is worth noting at this point that, thanks to permutation-based coding, it is not necessary to add any constraints to the model, such as those required when the TSP is defined using the QUBO or CQM formulations. This is undoubtedly an advantage for the \texttt{NL-Hybrid}.

In the following Python code snippet, we show how to initialize the model and the decision variables. We also show how to introduce the information needed to describe the problem.

\begin{python}
from dwave.optimization.model import Model

#Initializing the model entity
tsp_model = Model()

#Defining the variable as a list of size N
route = tsp_model.list(N)

#Entering the cost matrix as constant 
cost_matrix = tsp_model.constant(C)
\end{python}

\hltext{Regarding the objective function, which must be minimized, it can be mathematically formulated as follows, being $\mathbf{x}$ a \texttt{list} representing a feasible TSP \texttt{route}}:

\begin{equation}
    \begin{split}
    f(\mathbf{x}) = \sum_{i=1}^{N-1}cost\_matrix_{x_i,x_{i+1}}\\
    + cost\_matrix_{x_N,x_{1}}
    \end{split}
\end{equation}
\hltext{Which is represented in code as shown in the following snippet}:

\begin{python}
route_cost = cost_matrix[route[:-1], route[1:]]
return_cost = cost_matrix[route[-1], route[0]]

# Sum the costs of the full route. 
complete_cost = route_cost.sum()+return_cost.sum()
    
# The objective is introduced using model.minimize
tsp_model.minimize(complete_cost)

\end{python}

%-------------------------
\subsection{Knapsack Problem}
\label{sec:KP}
%-------------------------

In summary, KP consists of a set $I$ of $N$ items, describing each one by its weight ($w_i$) and its profit ($v_i$), which must be packed into a knapsack with a maximum capacity $C$. The objective is to choose a subset of items to be stored that maximizes the profit obtained and does not exceed $C$.

Like the TSP, the KP is a problem that benefits from the features of \texttt{NL-Hybrid}, since we can use the \texttt{set(number\_variables)} type to define the decision variables of the problem. Thus, \texttt{set(N)} represents a subset of $I$. We show the initialization process of the KP-related model in the following Python code snippet, where $W$ and $V$ are two sets that include the weights and benefits of all items, respectively.

\begin{python}
from dwave.optimization.model import Model

#Initializing the model entity
kp_model = Model()

#Defining the variable as a set of size N
items = kp_model.set(N)

#Entering problem information as constants
capacity = kp_model.constant(C)
weights = kp_model.constant(W)
profits = kp_model.constant(V)
\end{python}

In contrast to MCP and TSP, in the case of the KP it is necessary to add a restriction, which ensures that the items placed in the backpack do not exceed $C$.

\begin{python}
capacity_check = weights[items].sum() <= capacity

#Restrictions are introduced using 
#model.add_constraint method
model.add_constraint(capacity_check)
\end{python}

\hltext{Finally, the objective function, which must be maximized natively, can be mathematically formulated as follows, being $\mathbf{x}$ a feasible \texttt{set} of \texttt{items}, and $M$ the size of this \texttt{set}}:

\begin{equation}
    \begin{split}
    f(\mathbf{x}) = \sum_{i=1}^{M}values_{x_i}
    \end{split}
\end{equation}
Which can be \hltext{represented in code} as follows:

\begin{python}
# Sum the profits of introduced items. 
sum_values = profits[items].sum()
    
# The objective is introduced using model.minimize
kp_model.minimize(-sum_values)

\end{python}

\hltext{It is worth noting that \texttt{NL-Hybrid} method only allows for the minimization of the objective function. This is why problems natively designed to maximize an objective, such as the KP, are handled by adding a negative sign to the function, as seen in the code above. This situation also occurs with the MCP.}

%-------------------------
\subsection{Maximum Cut Problem}
\label{sec:MCP}
%-------------------------

Taking into account a directed graph $G$ made up of $N$ nodes and a weight matrix $W$ of size $NxN$, the objective of the MCP is to divide $N$ into two subsets such that the sum of the weights of the cut edges is maximum.

In order to formulate the MCP to be solved by \texttt{NL-Hybrid}, the set $X={x_1,\dots,x_N}$ of binary decision variables has been defined, where $x_i$ is 0 if node $i$ is part of the first subset and 1 otherwise.

We show the initialization process of the model in the following Python code snippet.

\begin{python}
from dwave.optimization.model import Model

#Initializing the model entity
mcp_model = Model()

#Defining the N bynary decision variables
nodes = mcp_model.binary(N)

#Entering the weights into the model as constant 
weights = mcp_model.constant(W)
\end{python}

\hltext{Furthermore, the objective function, which must be maximized natively, can be mathematically formulated as follows, being $\mathbf{x}$ a feasible binary solution to the MCP}:

\begin{equation}\label{eq:o_1}
    f(\mathbf{x}) = \sum_{i=1}^{N}\sum_{j=1}^N |x_i-x_j|*weights_{i,j}.
\end{equation}
\hltext{Which can be made explicit in code using the following this Python snippet}:

\begin{python}
obj = None

for i in range(N):
 for j in range(N):
  if i!=j:
   if obj is None:
    obj = abs(nodes[i]-nodes[j])*weights[i][j])

   else:
    obj = obj+(abs(nodes[i]-nodes[j]) 
    * weights[i][j]))

#The objective is introduced using model.minimize
mcp_model.minimize(-obj)

\end{python}

It should be noted that, because of the nature of the MCP, no problem constraints are needed.

%-------------------------
\subsection{Executing the problem}
\label{sec:exe}
%-------------------------

Once the problem has been modeled following the steps described in the previous subsections, it is ready to be submitted to \texttt{NL-Hybrid}. This process is carried out in the same way as the rest of the HSS portfolio, as can be seen in the following snippet:

\begin{python}
from dwave.system import LeapHybridNLSampler

sapi_token = 'XXXXX'
dwave_url = 'https://cloud.dwavesys.com/sapi'

sampler = LeapHybridNLSampler(token=sapi_token,
                              endpoint=dwave_url)
sampler.sample(model)
\end{python}

Finally, the results-collecting process differs from the rest of the HSS solvers. In this case, \texttt{NL-Hybrid} deposits the complete set of outcomes in the \textit{Model} itself. The results, as well as their energies, can be read as follows:

\begin{python}
from dwave.system import LeapHybridNLSampler
from dwave.optimization.model import Model

for i in range(model.states.size()):
    solution = next(model.iter_decisions()).
                    state(i).astype(int)
    solution_energy = model.objective.state(i)
\end{python}

\hltext{The code above produce an outcome as we represent in the following code snippet, centered on a TSP example composed of 9 nodes.}
\begin{python}
solution - objective function
[7 8 6 4 2 3 0 1 5] - 2134.0
[1 5 7 8 6 4 2 3 0] - 2134.0
[1 0 3 2 4 6 8 7 5] - 2134.0
[4 6 8 7 5 1 0 3 2] - 2134.0
[3 2 4 6 8 7 5 1 0] - 2134.0
...
\end{python}

%-------------------------
\section{Experimentation and Results}
\label{sec:expres}
%-------------------------

The main elements of the experiments carried out are described in this section. First, we detail the characteristics of the benchmark used in Section \ref{sec:benchmark}. Then, in Section \ref{sec:setting}, we describe the design choices adopted. Lastly, we show and analyze the results obtained in Section \ref{sec:results}.

%-------------------------
\subsection{Benchmark Description}
\label{sec:benchmark}
%-------------------------

\begin{table*}[t]
  \caption{Summary of the TSP, KP and MCP instances used}
  \label{tab:dataset}
    \centering
    \resizebox{1.8\columnwidth}{!}{
        \begin{tabular}{cc|cc|cc}
            \toprule
            \multicolumn{2}{c|}{\bf Traveling Salesman Problem} & \multicolumn{2}{c|}{\bf Knapsack Problem}  & \multicolumn{2}{c}{\bf Maximum Cut Problem} \\
            \textbf{Instance} & Description & \textbf{Instance} & Description & \textbf{Instance} & Description\\
            \midrule
            \textbf{wi7} & 7-noded complete graph & \textbf{s50\_1 } & 50 packages and a capacity of 11793 & \textbf{MC\_10} & 10-noded graph with 37 edges\\
            \textbf{dj8} & 8-noded complete graph & \textbf{s50\_2 } & 50 packages and a capacity of 135607 & \textbf{MC\_20} & 30-noded graph with 148 edges\\
            \textbf{dj9} & 9-noded complete graph & \textbf{s100\_1 } & 100 packages and a capacity of 238613 & \textbf{MC\_40} & 40-noded graph with 570 edges\\
            \textbf{dj15} & 15-noded complete graph & \textbf{s100\_2 } & 100 packages and a capacity of 233898 & \textbf{MC\_50} & 50-noded graph with 1013 edges\\
            \textbf{dj22} & 22-noded complete graph & \textbf{s200\_1 } & 200 packages and a capacity of 53104 & \textbf{MC\_60} & 60-noded graph with 1450 edges\\
            \textbf{wi25} & 25-noded complete graph & \textbf{s200\_2 } & 200 packages and a capacity of 51353 & \textbf{MC\_80} & 80-noded graph with 2563 edges\\
            \textbf{wi29} & 29-noded complete graph & \textbf{s200\_3 } & 200 packages and a capacity of 462929 & \textbf{MC\_90} & 90-noded graph with 3200 edges\\
            \textbf{dj38} & 38-noded complete graph & \textbf{s500\_1 } & 500 packages and a capacity of 128354 & \textbf{MC\_100} & 100-noded graph with 3902 edges\\
            \textbf{eil51} & 51-noded complete graph & \textbf{s500\_2 } & 500 packages and a capacity of 119918 & \textbf{MC\_120} & 120-noded graph with 5696 edges\\
            \textbf{berlin52} & 52-noded complete graph & \textbf{s500\_3 } & 500 packages and a capacity of 1236838 & \textbf{MC\_140} & 140-noded graph with 7912 edges\\
            \textbf{st70} & 70-noded complete graph & \textbf{s1000\_1 } & 1000 packages and a capacity of 2494539 & \textbf{MC\_150} & 150-noded graph with 8822 edges\\
            \textbf{eil76} & 76-noded complete graph & \textbf{s1000\_2 } & 1000 packages and a capacity of 243197 & \textbf{MC\_170} & 170-noded graph with 11486 edges\\
            \textbf{rat99} & 99-noded complete graph & \textbf{s1000\_3 } & 1000 packages and a capacity of 2429393 & \textbf{MC\_200} & 200-noded graph with 15916 edges\\
            \textbf{eil101} & 101-noded complete graph & \textbf{s2000\_1 } & 2000 packages and a capacity of 498452 & \textbf{MC\_220} & 220-noded graph with 19423 edges\\
            \textbf{pr107} & 107-noded complete graph & \textbf{s2000\_2 } & 2000 packages and a capacity of 497847 & \textbf{MC\_250} & 250-noded graph with 24831 edges\\

            \bottomrule
        \end{tabular}
    }
\end{table*}

As mentioned, a benchmark composed of 45 instances has been used, equally distributed over three combinatorial optimization problems: 

\begin{enumerate}
    \item \textbf{TSP}, for which instances of sizes between 7 and 107 nodes have been used. %The size of each dataset is explicitly represented in its name. 
    Instances with a size equal to or less than 25 nodes have been obtained from the QC-oriented benchmark QOPTLib \cite{osaba2023qoptlib}, while the rest have been obtained from the well-known TSPLib \cite{reinelt1991tsplib}. %For the TSP, the objective function represents the total distance of the calculated route.
    \item \textbf{KP}, for which each case is named \texttt{sX\_Y}, where \texttt{X} is the number of items and \texttt{Y} is a suffix to distinguish the set of instances with the same \texttt{X}. All instances have been obtained from the KPLib benchmark\footnote{https://github.com/likr/kplib}, described in \cite{kellerer2004exact}. %For this problem, the objective function is calculated by aggregating the sum of the values associated with the items in the backpack.
    \item \textbf{MCP}, for which each instance is coined \texttt{MC\_X}, with \texttt{X} being the number of nodes that define the graph. Ten of the instances have been obtained from the QOPTLib mentioned above, while graphs of sizes 80, 90, 120, 140, and 170 have been generated ad hoc for this study. %The objective function of this problem represents the sum of the weights of the cut edges.
\end{enumerate}

In Table \ref{tab:dataset}, we summarize the main characteristics of each instance considered. Finally, to improve the reproducibility of this study, all 45 instances are openly available in \cite{PaperRep}. Furthermore, the newly created MCP instances have been generated randomly using a Python script, which is also available in the same repository.

%-------------------------
\subsection{Experimental Setting}
\label{sec:setting}
%-------------------------

Using the benchmark described above, the main objective is to analyze the performance of the \texttt{NL-Hybrid}, and to compare the results obtained by this method with those obtained by three D-Wave-based counterparts: \hltext{the QPU}, \texttt{BQM-Hybrid} and \texttt{CQM-Hybrid}. Our main motivation for choosing these counterparts is to measure the main contribution that \texttt{NL-Hybrid} provides in relation to previously available methods. \hltext{More specifically, the \texttt{BQM-Hybrid} has been employed in these tests because it is the most widely employed HSS solver and its input format is similar to that of the QPU. Additionally, the \texttt{CQM-Hybrid} has been used because it is the most advanced algorithm in the portfolio and is easy to use when modeling the problems addressed. Finally, the \texttt{DQM-Hybrid} has been set aside due to the little attention it has received in the literature and the complexity of its use compared to the other solvers within HSS.}

\begin{figure*}[t]
    \centering
    \includegraphics[width=0.90\linewidth]{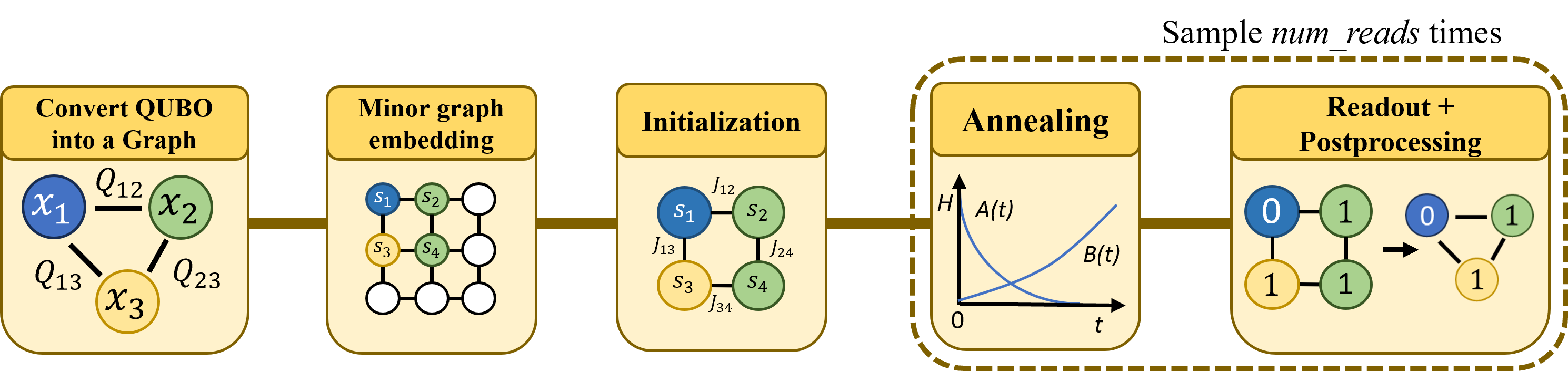}
    \caption{Graphical representation of the quantum annealing process followed by the D-Wave's QPU used in this work}
    \label{fig:QPU}
\end{figure*}

\hltext{Regarding the QPU, the \texttt{Advantage\_system6.4} device has been used, which is the most recent from D-Wave at the time this work was written. This computer features 5,616 qubits and more than 35,000 couplers arranged in a Pegasus topology. Similar to hybrid solvers, the QPU has been accessed through the \texttt{D-Wave} \texttt{Leap} cloud service, and the common forward annealing process has been executed, which consists of the following main steps \cite{yarkoni2022quantum}: \textit{i}) convert QUBO into a graph, \textit{ii}) minor graph embedding, \textit{iii}) QPU initialization, \textit{iv}) annealing, \textit{v}) readout, and \textit{vi}) postprocessing. In Figure \ref{fig:QPU}, we graphically represent this process to facilitate its understanding.}

In relation to the parameterization, the default values have been used for all the solvers for the sake of fairness. For hybrid solver versions, v2.2, v1.12, and v1.1 have been used for \texttt{BQM-Hybrid}, \texttt{CQM-Hybrid}, and \texttt{NL-Hybrid}, respectively. For the QPU, the default parameters have also been used.

\hltext{Finally, \textit{Qiskit\_Optimization}\footnote{\url{https://qiskit.org/ecosystem/optimization/apidocs/qiskit_optimization.applications.html}} v0.6.1 open libraries have been employed to aid in solving the three optimization problems using the QPU and \texttt{BQM-Hybrid}. These libraries have been utilized because they enable tasks such as data reading, solution decoding, result evaluation, and, most importantly, the automatic construction of the QUBOs. More specifically, to build the QUBO formulations, these classes relax the constraints using a penalty model, with the coefficient being automatically estimated. Finally, it is worth noting that \textit{Qiskit\_Optimization} does not present any incompatibility with any versions of Qiskit, such as the current v1.2.}

The CQM implementations of the three problems have been developed ad hoc for this research. Finally, as mentioned above, the NL implementation of the MCP has been done ad hoc for this work, while the implementations of TSP and KP are slight adaptations of the open code published by D-Wave. To improve the reproducibility of this research, all codes are available from the corresponding author upon reasonable request.

%-------------------------
\subsection{Results}
\label{sec:results}
%-------------------------

\begin{figure}[t]
    \centering
    \includegraphics[width=1.0\linewidth]{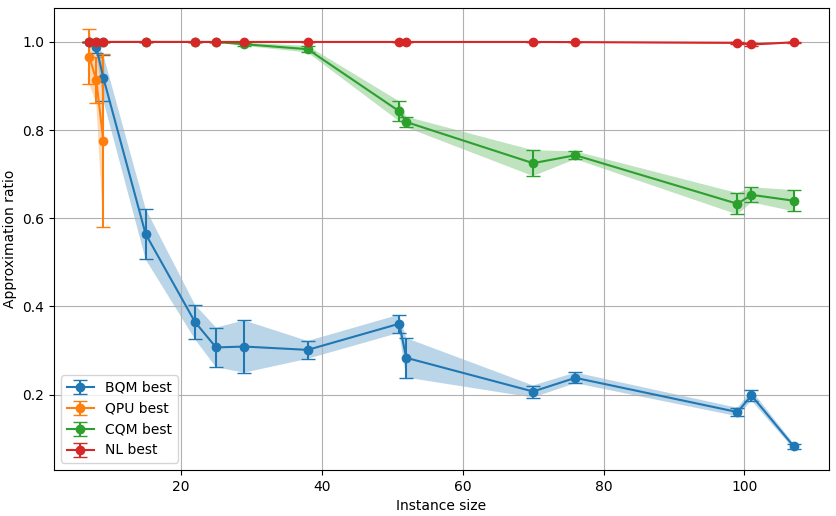}
    \caption{The average approximation ratio of the best solutions found by each method for the TSP instances.}
    \label{fig:ar_TSP}
\end{figure}

In order to obtain representative results, all the outcomes presented in this section have been obtained after 10 independent runs per method and instance. Regarding TSP, we show in Figure \ref{fig:ar_TSP} the average of the best solution reached by each solver. It should be noted that the metric used to represent the quality of the solutions is the approximation ratio. \hltext{This metric has been applied to all the methods executed, and it is particularly suitable for measuring the distance between the obtained outcomes and known optimal solutions in terms of the objective function cost. Thus, the approximation ratio is calculated by using the cost of an obtained solution and the optimal value of the problem being solved. The optimal values for each instance have been obtained from \cite{osaba2023qoptlib} and \cite{reinelt1991tsplib}}. 

\begin{figure}[t]
    \centering
    \includegraphics[width=1.0\linewidth]{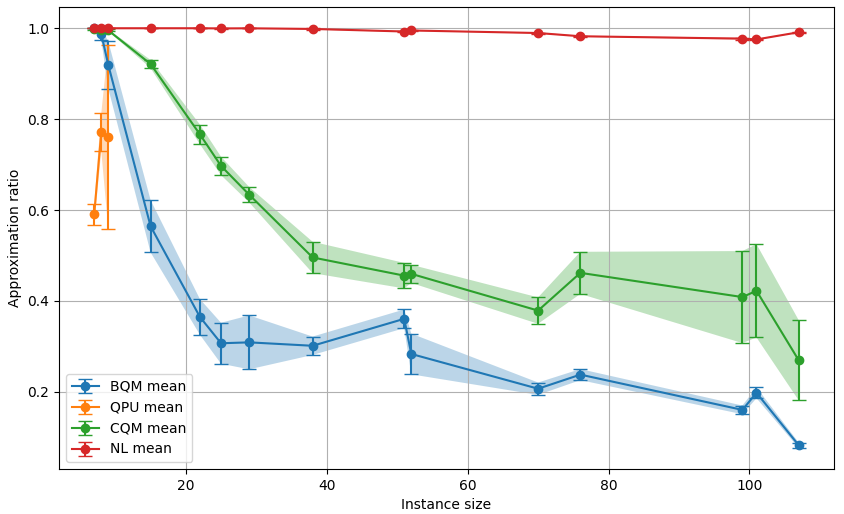}
    \caption{The average approximation ratio of the whole \textit{sample-sets} found by each method for the TSP instances.}
    \label{fig:ar_mean_TSP}
\end{figure}

In addition to the best solution found, each solver provides a set of solutions called \textit{sample-set} for each execution. It is possible that there are repeated solutions in this \textit{sample-set}. For this reason, the reliability of a method can be measured by analyzing the entire \textit{sample-set}, with it being preferable that the quality of the solution set approaches the optimal solution to the problem. Thus, to represent the robustness of each method, the averages of the complete \textit{sample-sets} obtained by each technique are presented in Figure \ref{fig:ar_mean_TSP}. 

\hltext{A simple glance at Figure \ref{fig:ar_TSP} and Figure \ref{fig:ar_mean_TSP} is enough to detect the superiority of the \texttt{NL-Hybrid} solver over its competitors.} To verify this improvement, two statistical tests have been performed with the results shown in these figures.

First, we applied Friedman's non-parametric test to determine if there are significant differences among \texttt{NL-Hybrid}, \texttt{CQM-Hybrid}, and \texttt{BQM-Hybrid}. It should be noted that QPU has been left out of all these tests as it does not provide outcomes for all datasets. The results of this test are presented in Table \ref{tab:results_friedman_TSP}. \hltext{Specifically, the average ranking value returned by the Friedman non-parametric test is presented for each of the compared algorithms. A lower rank indicates better performance.} Furthermore, on the left-hand side of Table \ref{tab:results_friedman_TSP}, we show the results referring to the best results obtained, while on the right-hand side, we show those concerning the complete \textit{sample-set} mean.

The Friedman statistics obtained in these tests are 26.53 and 28.13. With a 99\% confidence interval, the critical value in a $\chi^2$ distribution with 2 degrees of freedom is 9.21. Since both statistics are greater than this critical value, we can conclude that there are significant differences among the results, with \texttt{NL-Hybrid} having the lowest rank. 

Following the results described above, we conducted Holm's post-hoc test to evaluate the statistical significance of \texttt{NL-Hybrid}'s superior performance. The adjusted p-values from Holm's post-hoc procedure are presented in Table \ref{tab:results_holms_TSP}. Upon analyzing these results, and considering that all p-values are below 0.05, we can confidently conclude that \texttt{NL-Hybrid} significantly outperforms \texttt{BQM-Hybrid} and \texttt{CQM-Hybrid} with 95\% confidence.

\begin{table}[tbh]
	\centering
    \caption{Average rankings obtained using the Friedman's test for the TSP experimentation}
	\resizebox{1.0\columnwidth}{!}{
		\begin{tabular}{cc|cc}\hline
            \multicolumn{2}{c|}{\bf Best solution found} & \multicolumn{2}{c}{\bf Full \textit{sample-set} average} \\ 
			Algorithm&Average Ranking & Algorithm&Average Ranking\\\hline
			\texttt{NL-Hybrid}&1.3333 & \texttt{NL-Hybrid}&1\\
			\texttt{CQM-Hybrid}&1.8667 & \texttt{CQM-Hybrid}&2.0667\\
			\texttt{BQM-Hybrid}&2.8 & \texttt{BQM-Hybrid}&2.9333\\
			\hline
		\end{tabular}
	}
	\label{tab:results_friedman_TSP}
\end{table}

\begin{table}[tbh]
	\centering
    \caption{Results obtained using the Holm's post-hoc procedure for the TSP. \texttt{NL-Hybrid} used as control algorithm.}
	\resizebox{0.8\columnwidth}{!}{
		\begin{tabular}{cc|cc}\hline
        \multicolumn{2}{c|}{\bf Best solution found} & \multicolumn{2}{c}{\bf Full \textit{sample-set} average} \\ 
		Algorithm & Adjusted $p$ & Algorithm & Adjusted $p$\\
		\hline
		\texttt{CQM-Hybrid}&0.04461 & \texttt{CQM-Hybrid}&0.003487\\
		\texttt{BQM-Hybrid}&0.0000018 & \texttt{BQM-Hybrid}&0.000027\\\hline
		\end{tabular}
	}
	\label{tab:results_holms_TSP}
\end{table}

\begin{figure}[t]
    \centering
    \includegraphics[width=1.0\linewidth]{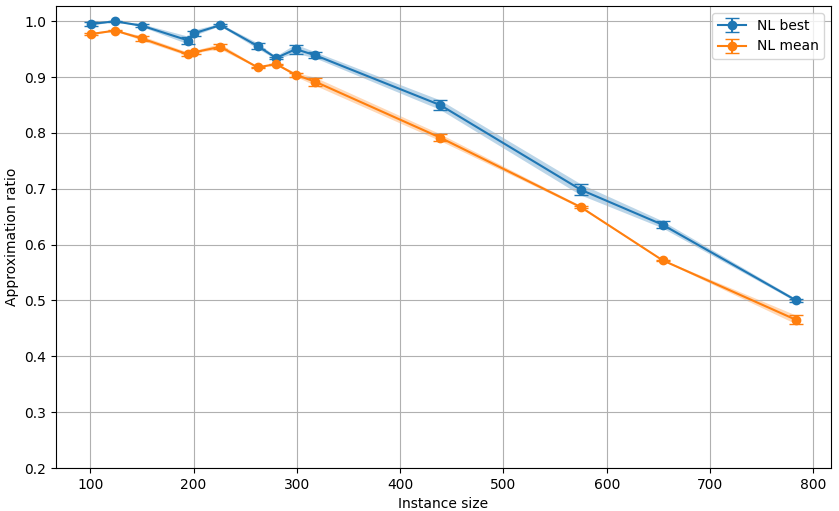}
    \caption{The average of best solutions and of the whole \textit{sample-sets} found by the \texttt{NL-Hybrid} for big TSP instances.}
    \label{fig:big_TSP}
\end{figure}

Finally, as can be seen in Figure \ref{fig:ar_TSP}, the performance of \texttt{NL-Hybrid} in all the datasets considered is almost perfect. For this reason, and to get a glimpse of the limits of \texttt{NL-Hybrid}'s performance, we have conducted additional experiments with TSPLib instances composed of between 100 and 783 nodes. We show the average of the best results and the average of the complete \textit{sample-sets} in Figure \ref{fig:big_TSP}, where it can be seen how \texttt{NL-Hybrid} obtains remarkable results (above 0.8 with respect to the optimum) in instances of up to 439 nodes. This performance is a substantial improvement over other hybrid methods in the literature.

\begin{figure}[t]
    \centering
    \includegraphics[width=1.0\linewidth]{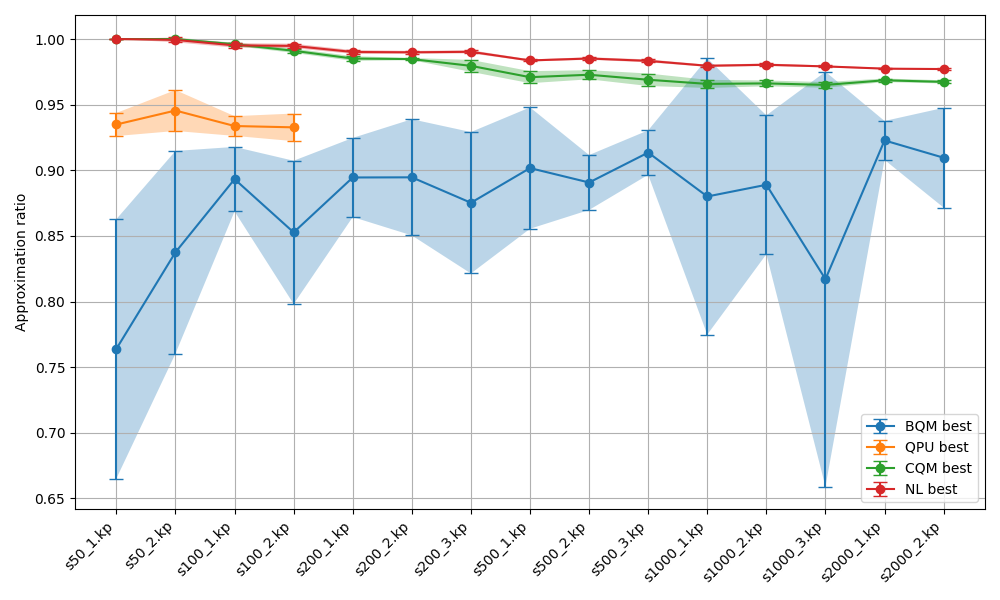}
    \caption{The average approximation ratio of the best solutions found by each method for the KP instances.}
    \label{fig:ar_KP}
\end{figure}

\begin{figure}[t]
    \centering
    \includegraphics[width=1.0\linewidth]{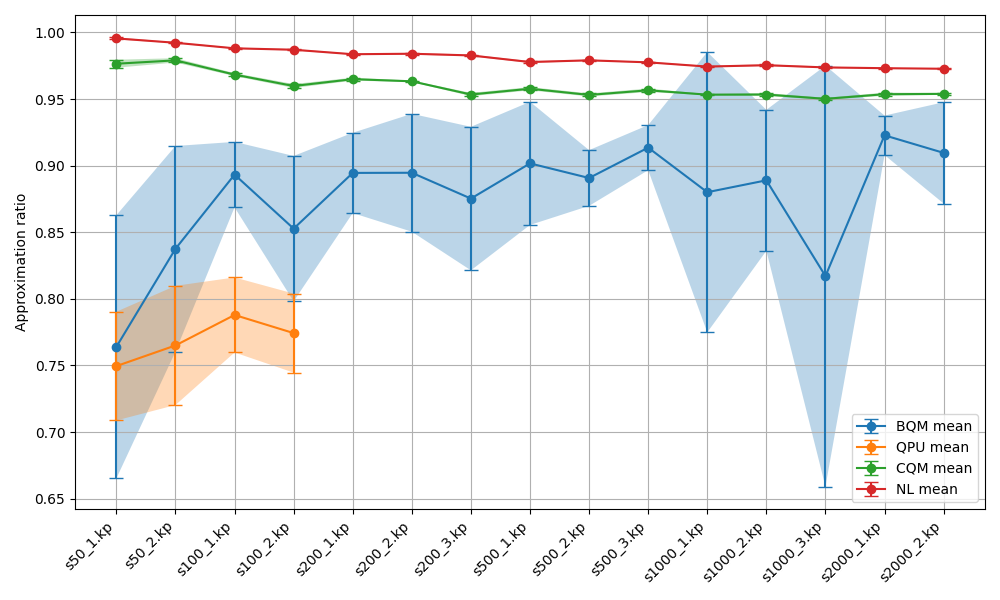}
    \caption{The average approximation ratio of the whole \textit{sample-sets} found by each method for the KP instances.}
    \label{fig:ar_mean_KP}
\end{figure}

Similar conclusions can be drawn if we focus our attention on the tests carried out with the KP. For the KP, we show in Figure \ref{fig:ar_KP} and Figure \ref{fig:ar_mean_KP} the results obtained in terms of the best solution found per run and the average of the entire \textit{sample-sets}, respectively. In this case, the optimal values used as the baseline have been obtained by solving each instance through \texttt{Google OR-Tools}. As mentioned, \texttt{NL-Hybrid} proves to be superior to its competitors also for the KP. The outcomes obtained after the execution of both Friedman's and Holm's post hoc tests are depicted in Table \ref{tab:results_friedman_KP} and Table \ref{tab:results_holms_KP}, respectively. On the one hand, the Friedman statistics obtained are 26.53 and 30. Given that both statistics exceed the critical value, we can infer that there are significant differences between the results, with \texttt{NL-Hybrid} achieving the lowest rank. On the other hand, because all p-values are below 0.05, we can conclude that Holm's post-hoc test supports the conclusion that \texttt{NL-Hybrid} is significantly better than \texttt{CQM-Hybrid} and \texttt{BQM-Hybrid} with 95\% confidence.

\begin{table}[tbh]
	\centering
    \caption{Average rankings obtained using the Friedman's test for the KP experimentation}
	\resizebox{1.0\columnwidth}{!}{
		\begin{tabular}{cc|cc}\hline
            \multicolumn{2}{c|}{\bf Best solution found} & \multicolumn{2}{c}{\bf Full \textit{sample-set} average} \\ 
			Algorithm&Average Ranking & Algorithm&Average Ranking\\\hline
			\texttt{NL-Hybrid}&1.1333 & \texttt{NL-Hybrid}&1\\
			\texttt{CQM-Hybrid}&1.8667 & \texttt{CQM-Hybrid}&2\\
			\texttt{BQM-Hybrid}&3 & \texttt{BQM-Hybrid}&3\\
			\hline
		\end{tabular}
	}
	\label{tab:results_friedman_KP}
\end{table}

\begin{table}[tbh]
	\centering
    \caption{Results obtained using the Holm's post-hoc procedure for the KP. \texttt{NL-Hybrid} used as control algorithm.}
	\resizebox{0.8\columnwidth}{!}{
		\begin{tabular}{cc|cc}\hline
        \multicolumn{2}{c|}{\bf Best solution found} & \multicolumn{2}{c}{\bf Full \textit{sample-set} average} \\ 
		Algorithm & Adjusted $p$ & Algorithm & Adjusted $p$\\
		\hline
		\texttt{CQM-Hybrid}&0.04461 & \texttt{CQM-Hybrid}&0.00617\\
		\texttt{BQM-Hybrid}&0.000001 & \texttt{BQM-Hybrid}&0\\\hline
		\end{tabular}
	}
	\label{tab:results_holms_KP}
\end{table}

\begin{figure}[t]
    \centering
    \includegraphics[width=0.9\linewidth]{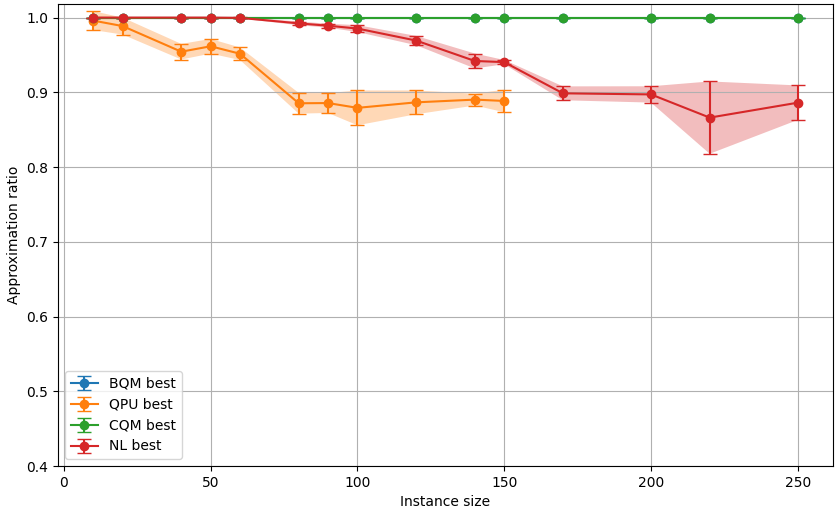}
    \caption{The average approximation ratio of the best solutions found by each method for the MCP instances.}
    \label{fig:ar_MCP}
\end{figure}

\begin{figure}[t]
    \centering
    \includegraphics[width=1.0\linewidth]{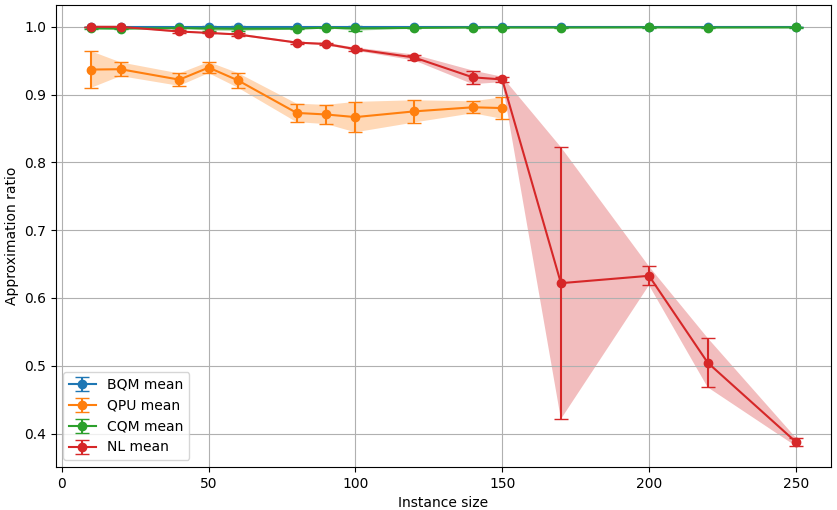}
    \caption{The average approximation ratio of the whole \textit{sample-sets} found by each method for the MCP instances.}
    \label{fig:ar_mean_MCP}
\end{figure}

Finally, Figures \ref{fig:ar_MCP} and \ref{fig:ar_mean_MCP} present the results for MCP, where the conclusions differ significantly from the previous ones. %In this case, the optimum values of the 15 instances have been obtained using the industry-oriented Quantagonia's Hybrid Solver\footnote{\url{https://www.quantagonia.com/hybridsolver}}. 
A glance at the results shows that \texttt{NL-Hybrid} only outperforms the QPU, offering considerably lower performance compared to \texttt{BQM-Hybrid} and \texttt{CQM-Hybrid}, both of which demonstrate outstanding suitability for this problem. To assess if the differences between \texttt{NL-Hybrid} and the other algorithms are statistically significant, we used the Wilcoxon rank-sum test. \hltext{The results are included in Table \ref{Table:Wilcoxon}}. Each cell displays the two metrics considered (best solution found and average of the sample set) using one of the following symbols: ``$\blacktriangle$'' indicates that \texttt{NL-Hybrid} has produced better results than the algorithm in the column with 99\% confidence, and ``$\triangledown$'' denotes that the algorithm in the column is statistically superior to \texttt{NL-Hybrid}.

\begin{table}[h!]
	\centering
    \caption{Wilcoxon test results. Each cell contains a symbol per metric (best solution found and average of the \textit{sample-set}).}
	\resizebox{0.8\columnwidth}{!}{\begin{tabular}{
				 l | c | c | c 
			}
			\hline & QPU & \texttt{BQM-Hybrid} & \texttt{CQM-Hybrid}\\
			\hline 
			\texttt{NL-Hybrid}	& $\blacktriangle$ $\blacktriangle$ &  $\triangledown$ $\triangledown$ & $\triangledown$ $\triangledown$\\ 
			\hline
	\end{tabular}}
    \label{Table:Wilcoxon}
\end{table}

\hltext{Thus, as can be observed in Table \ref{Table:Wilcoxon}}, \texttt{NL-Hybrid} significantly outperforms QPU in both metrics, but the superiority of \texttt{BQM-Hybrid} and \texttt{CQM-Hybrid} is also statistically significant for both metrics. \hltext{However, this performance does not detract from the value of \texttt{NL-Hybrid}}. Although these results seem counterintuitive compared to those of TSP and KP, they are consistent with what was expressed in Section \ref{sec:desc}. That is, depending on the type of problem and the coding of its decision variables, \texttt{NL-Hybrid} may not always be the most suitable algorithm. Specifically, as shown through this experimentation and supported by D-Wave \cite{D-Wave}, for problems primarily involving binary variables or integers with low ranges, \texttt{CQM-Hybrid} or \texttt{BQM-Hybrid} are more appropriate. This situation opens up a wide range of future work to be carried out around the \texttt{NL-Solver} and its applicability to industrial problems, which we detail in the following section.

%-------------------------
\section{Conclusions and Further Work}
\label{sec:conclusion}
%-------------------------

This paper focuses on the recently introduced \texttt{NL-Hybrid} solver, evaluating its performance through a benchmark of 45 instances across three combinatorial optimization problems. The results have been compared with three different counterparts. Additionally, to facilitate the use of this relatively unexplored solver, we have provided detailed implementation guidelines for solving the three optimization problems considered.

\hltext{Given the results obtained, it is prudent to conclude that D-Wave's \texttt{NL-Hybrid} emerges as a promising alternative in the realm of hybrid classical-quantum algorithms. The \texttt{NL-Hybrid} has demonstrated superior performance compared to its competitors, particularly in constrained problems where the new decision variables are applicable. This algorithm proves to be a high-quality approach for such complex problems, efficiently solving large instances and achieving near-optimal solutions. It is likely that \texttt{NL-Hybrid} will become a flagship in the field, primarily due to the innovative types of variables it incorporates, which enable the efficient handling of highly complex problems that have posed significant challenges to existing hybrid solvers.}

Furthermore, an important factor to consider when evaluating \texttt{NL-Hybrid} is its ease of use. As shown in Section \ref{sec:implementation}, the way variables, constraints, and the objective function are defined is very intuitive. This is not a trivial matter, as it helps to bring and expand the use of quantum computing to a larger group of researchers who may not be familiar with complex quantum concepts.

\hltext{However, not everything that glitters is gold, as \texttt{NL-Hybrid} presents certain limitations that warrant more detailed exploration. Investigating these weaknesses will enable the scientific community to delineate the boundaries of this solver and determine the contexts in which it can perform optimally. Indeed, the \texttt{NL-Hybrid} has shown inferior performance compared to other hybrid alternatives in the MCP, where the decision variables used were of the binary type and the problem had no constrains. Although this is a limitation, it is not unexpected. In fact, unconstrained problems are ideally suited for the other methods, since there is no need for penalty models. This observation aligns with the statements previously made by D-Wave in \cite{D-Wave}.}

%In summary, it is a mistake to consider \texttt{NL-Hybrid} as a general-purpose solver, as the goal behind its implementation is not to replace the existing algorithms in \texttt{HSS}, but to complement them. In this way, \texttt{NL-Hybrid} is specifically designed to solve those problems that can take advantage of the new types of decision variables.

\hltext{Also, like any hybrid algorithm, there are further limitations that require further investigation. One of these limitations regards the overhead of asynchronous communications. Since quantum and classical computers are separate physical devices, each with its own computational interface and data transfer pipeline, the \texttt{NL-Hybrid} is constrained to repeatedly switch contexts and exchange intermediate data between devices. This latency prevents classical decisions from influencing the evolution of the quantum state before qubits undergo decoherence. Another aspect that requires further investigation is the scalability of the algorithm when dealing with increasingly larger problems.}

With all this, the findings reported in this work have enabled us to identify a set of inspiring opportunities, paving the way for several future research directions. These are some of the most intriguing challenges to pursue:

\begin{itemize} 
    \item Analyze the performance of \texttt{NL-Hybrid} against other well-known optimization problems, such as Bin Packing, or the Job-shop Scheduling Problem. For this, it will be necessary to work on adapting the formulation of these problems to the new types of decision variables. 
    
    \item Study the performance of \texttt{NL-Hybrid} compared to other commercial hybrid methods, not only in terms of quality, but also in terms of scalability.
    
    \item Considering the \texttt{NL-Hybrid}'s ability to handle complex constraints, efforts will be made to solve problems composed of multiple constraints. The Rich-Vehicle Routing Problem or the Three-Dimensional Bin Packing Problem are suitable cases to advance in this line. 
    
    \item Explore alternative problem formulations to improve the performance of the \texttt{NL-Hybrid}. An example of this line of research is the Maximum Cut problem, where a formulation that allows the use of the new types of decision variables will be studied. 

\end{itemize}

%Undoubtedly, NL represents a new step forward in the implementation of hybrid algorithms, expanding the application of D-Wave's quantum annealers to problems with a clear industrial orientation. With all this, we encourage the rest of the scientific community to experiment and, in this way, establish the strengths and limits of this novel algorithm.

% ---- Bibliography ----
\bibliographystyle{ieeetr}
\bibliography{bibliography.bib}

% ---- Biographies ----
\begin{IEEEbiography}[{\includegraphics[width=1in,height=1.25in,clip,keepaspectratio]{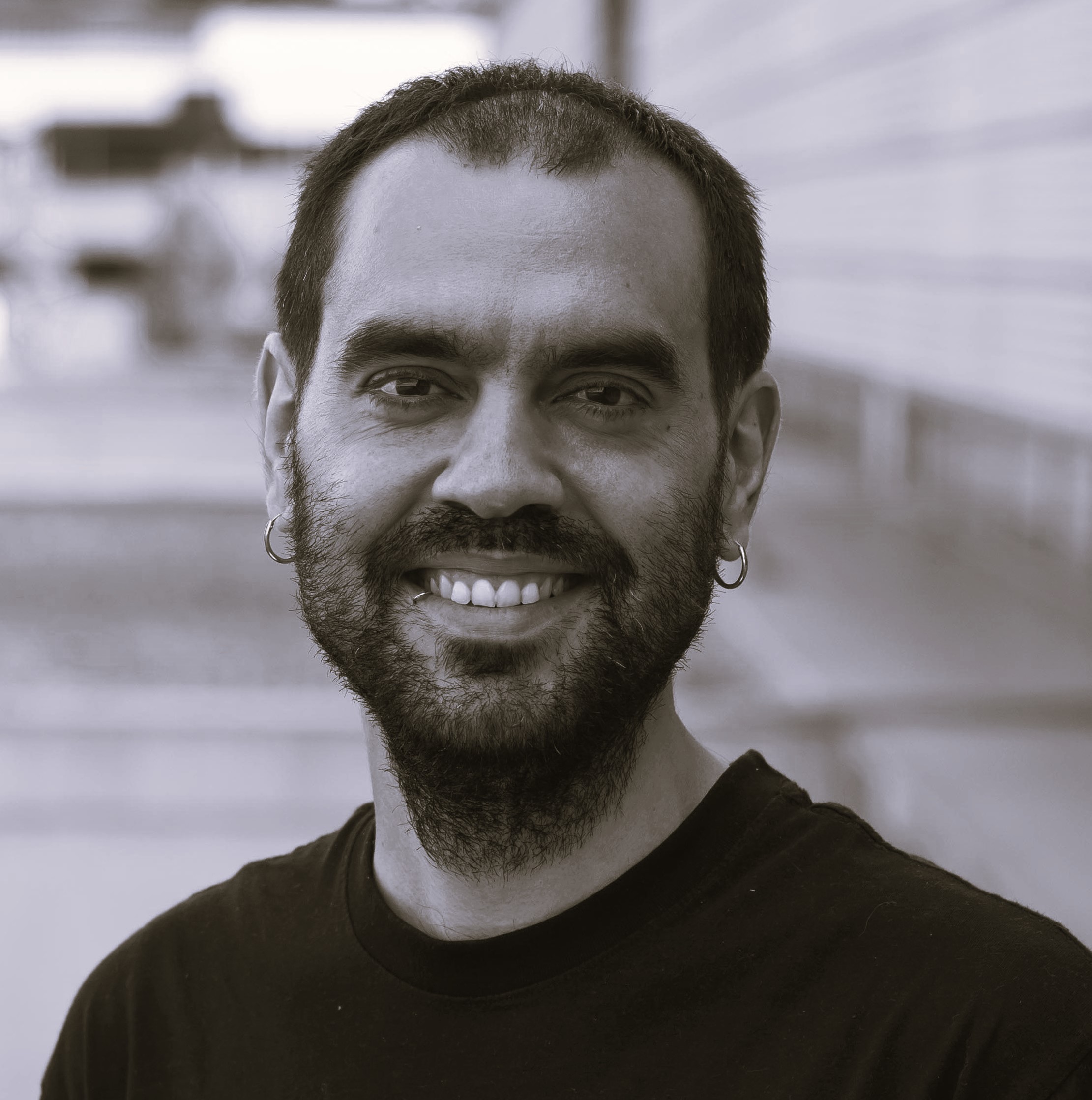}}]{Dr. Eneko Osaba} works at TECNALIA as principal researcher in the DIGITAL/Next area. He obtained his Ph.D. degree on Artificial Intelligence in 2015. He has participated in more than 30 research projects. He has contributed in the development of more than 170 papers, including more than 30 Q1. He has performed several stays in universities of United Kingdom, Italy and Malta. He has served as a member of the program and/or organizing committee in more than 60 international conferences. He is member of the editorial board of Data in Brief and Journal of Advanced Transportation. He has acted as guest editor in journals such as Neurocomputing, Journal of Supercomputing, Swarm and Evolutionary Computation and IEEE ITS Magazine. Finally, in 2022, Eneko was recognized by the Basque Research and Technology Alliance as one of the most promising young researchers of the Basque Country, Spain.
\end{IEEEbiography}

\begin{IEEEbiography}[{\includegraphics[width=1in,height=1.25in,clip,keepaspectratio]{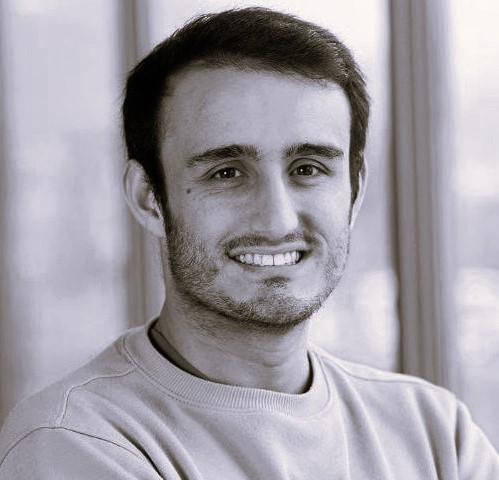}}]{Pablo Miranda-Rodriguez} works at TECNALIA as junior researcher in the DIGITAL/Next area. In 2019, he obtained a Bachelors degree in Physics and since 2020, he has a MSc in Mathematical Physics, both degrees with a specialization in black hole thermodynamics. In 2022, Pablo obtained a MSc in Quantum Computing, specializing in annealing methods for combinatorial optimization in finance. Since then, he has worked in hybrid computing, especially using the quantum annealing paradigm on optimization problems. He has participated in several proof-of-concept projects in the field of quantum computing.
\end{IEEEbiography}

\EOD

\end{document}